\documentclass[aps,pra,twocolumn,showpacs,floatfix]{revtex4-1}
\usepackage{graphicx,amsfonts,amssymb,amsmath,hyperref}
\usepackage{psfrag}
\usepackage{subfigure}

\newif\ifhyper
\hypertrue
\ifhyper
\hypersetup{
  citecolor = {green},
  colorlinks = {true}, 
  urlcolor = {blue} 
}

\begin{document}


\newcommand{\ie}{\textit{i.e.} }
\newcommand{\bea}{\begin{eqnarray}}
\newcommand{\eea}{\end{eqnarray}}
\newcommand{\beq}{\begin{equation}}
\newcommand{\eeq}{\end{equation}}
\newcommand{\benu}{\begin{enumerate}}
\newcommand{\enu}{\end{enumerate}}

\newcommand{\nb}{{\overline{n}}}
\newcommand{\cb}{{\overline{c}}}
\newcommand{\ka}{\kappa}

\newcommand{\kg}{k_G}
\newcommand{\kh}{k_h}
\newcommand{\khb}{k^B_h}
\newcommand{\lh}{l_h}
\newcommand{\lG}{l_G}
\newcommand{\ns}{n_s}
\newcommand{\ek}{\epsilon_k}
\newcommand{\eq}{\epsilon_{\bold{q}}}
\newcommand{\tq}{t_{\bold{q}}}
\newcommand{\gk}{\Gamma_k}
\newcommand{\gag}{\Gamma^{(2)}}
\newcommand{\rk}{R_k}
\newcommand{\dtr}{\tilde{\partial}_t }
\newcommand{\om}{\omega}
\newcommand{\intq}{\int_{q}}
\newcommand{\ga}{\Gamma_A}
\newcommand{\gb}{\Gamma_B} 
\newcommand{\gc}{\Gamma_C}
\newcommand{\intt}{\int_0^{\beta}{d\tau\,}}
\newcommand{\dt}{\partial_{\tau}}
\newcommand{\vq}{\bold{q}}
\def\den{\bar{n}}
\def\half{\frac{1}{2}}
\def\L{\Lambda}
\newcommand{\Tr}{{\rm Tr}} 
\newcommand{\tr}{{\rm tr}} 
\newcommand{\mean}[1]{\langle #1 \rangle}
\def\eps{\epsilon}
\def\gam{\gamma} 
\def\half{\frac{1}{2}}
\def\p{{\bf p}} 
\def\q{{\bf q}}
\def\r{{\bf r}}
\def\t{{\bf t}}
\def\u{{\bf u}}
\def\v{{\bf v}}
\def\x{{\bf x}}
\def\y{{\bf y}} 
\def\z{{\bf z}} 
\def\A{{\bf A}}
\def\B{{\bf B}}
\def\D{{\bf D}} 
\def\E{{\bf E}} 
\def\F{{\bf F}} 
\def\H{{\bf H}}  
\def\J{{\bf J}}
\def\K{{\bf K}} 

\def\L{{\bf L}}
\def\M{{\bf M}}  
\def\O{{\bf O}} 
\def\P{{\bf P}} 
\def\Q{{\bf Q}} 
\def\R{{\bf R}}
\def\S{{\bf S}}
\def\nablabf{\boldsymbol{\nabla}}

\def\w{\omega}
\def\wn{\omega_n}
\def\wnu{\omega_\nu}
\def\wp{\omega_p} 
\def\dmu{{\partial_\mu}}
\def\dl{{\partial_l}}  
\def\dt{\partial_t}
\def\dx{\partial_x}
\def\dy{\partial_y} 
\def\dtau{{\partial_\tau}}  
\def\det{{\rm det}} 

\def\dsum{\displaystyle \sum}
\def\dint{\displaystyle \int} 
\def\intt{\int_{-\infty}^\infty dt} 
\def\inttp{\int_{-\infty}^\infty dt'} 
\def\intk{\int_{\bf k}} 
\def\intkd{\int \frac{d^dk}{(2\pi)^d}}
\def\intq{\int_{\bf q}} 
\def\intr{\int d^dr}  
\def\dintr{\displaystyle \int d^dr} 
\def\intrp{\int d^dr'}
\def\dinttau{\displaystyle \int_0^\beta d\tau}
\def\dinttaup{\displaystyle \int_0^\beta d\tau'}
\def\inttau{\int_0^\beta d\tau}
\def\inttaup{\int_0^\beta d\tau'}
\def\intx{\int d^{d+1}x} 
\def\inttaur{\int_0^\beta d\tau \int d^dr}
\def\intinf{\int_{-\infty}^\infty}
\def\dinttaur{\displaystyle \int_0^\beta d\tau \int d^dr}
\def\dintinf{\displaystyle \int_{-\infty}^\infty}
\def\intw{\int_{-\infty}^\infty \frac{d\w}{2\pi}}
\def\calO{{\cal O}}


\title{Non-perturbative renormalization-group approach to the Bose-Hubbard model} 

\author{A. Ran\c{c}on and  N. Dupuis}
\affiliation{
 Laboratoire de Physique Th\'eorique de la Mati\`ere Condens\'ee, 
CNRS UMR 7600, \\ Universit\'e Pierre et Marie Curie, 4 Place Jussieu, 
75252 Paris Cedex 05,  France}

\date{February 17, 2010}

\begin{abstract} 
We present a non-perturbative renormalization-group approach to the Bose-Hubbard model. By taking as initial condition of the RG flow the (local) limit of decoupled sites, we take into account both local and long-distance fluctuations in a nontrivial way. This approach yields a phase diagram in very good quantitative agreement with the quantum Monte Carlo results and reproduces the two universality classes of the superfluid--Mott-insulator transition with a good estimate of the critical exponents. Furthermore, it reveals the crucial role of the ``Ginzburg length'' as a crossover length between a weakly- and a strongly-correlated superfluid phase.  
\end{abstract}
\maketitle

{\it Introduction.} In the last two decades, the non-perturbative renormalization-group (NPRG) approach has been successfully applied to many areas of physics, from high-energy physics to statistical and condensed-matter physics (for reviews, see Refs.~\cite{Berges02,Delamotte07}). Although the RG is often seen as a powerful tool to study the low-energy long-distance physics in the framework of effective field theories, it has recently been shown that the NPRG also applies to lattice models and enables to compute not only universal quantities (critical exponents) but also non-universal quantities (such as phase diagrams, transition temperatures, order parameters) which strongly depend on the microscopic parameters of the model (lattice type, strength of the interactions, etc.). This implementation of the NPRG is referred to as the lattice NPRG~\cite{Machado10}. 

In this paper, we report a NPRG study of the Bose-Hubbard model~\cite{Fisher89}. This approach yields a description of the superfluid--Mott-insulator transition which takes into account both local fluctuations (which drive the Mott transition and determine the phase diagram) and critical fluctuations in a nontrivial way. By comparing with the numerically exact lattice quantum Monte Carlo simulation (QMC), we show that the NPRG yields remarkably accurate results for the phase diagram. Moreover, contrary to the QMC simulation, we obtain the critical behavior at the Mott transition and recover the existence of two universality classes~\cite{Fisher89}. We also emphasize the crucial role of the Ginzburg length $\xi_G$ as a crossover length between a weakly- and a strongly-correlated superfluid phase.

{\it The non-perturbative RG.} The $d$-dimensional Bose-Hubbard model is defined by the (Euclidean) action
\begin{align}
S = \inttau \biggl\lbrace & \sum_\r \biggl[ \psi_\r^* (\dtau-\mu)\psi_\r + \frac{U}{2} (\psi_\r^*\psi_\r)^2 \biggr] \nonumber \\ & - t \sum_{\mean{\r,\r'}} \left(\psi_\r^* \psi_{\r'}+\mbox{c.c.}\right) \biggr\rbrace ,
\label{action}
\end{align}
where $\psi_\r(\tau)$ is a complex field and $\tau\in [0,\beta]$ an imaginary time with $\beta\to\infty$ the inverse temperature. $\lbrace\r\rbrace$ denotes the $N$ sites of the lattice and $\mean{\r,\r'}$ nearest-neighbor sites. $U$ is the on-site repulsion, $t$ the hopping amplitude and $\mu$ the chemical potential. (We take $\hbar=k_B=1$ throughout the paper.) 

The strategy of the NPRG is to build a family of models with action $S_k=S+\Delta S_k$ indexed by a momentum scale $k$ varying from a microscopic scale $\Lambda$ down to 0. This is achieved by adding to the action~(\ref{action}) the term $\Delta S_k=\inttau \sum_\q \psi^*_\q R_k(\q) \psi_\q$ ($\psi_\q$ is the Fourier transform of $\psi_\r$), where 
\beq 
R_k(\q) = - Z_{A,k} tk^2 \mbox{sgn}(t_\q) (1-y_\q)\Theta(1-y_\q) , 
\eeq
with $t_\q=-2t\sum_{i=1}^d \cos q_i$, $y_\q=(2dt-|t_\q|)/tk^2$ and $\Theta(x)$ the step function (we take the lattice spacing as the unit length). The $k$-dependent constant $Z_{A,k}$ is defined below. Since $R_{k=0}(\q)=0$, the action $S_{k=0}$ coincides with the action~(\ref{action}). On the other hand, for $k=\Lambda=\sqrt{2d}$, $R_\Lambda(\q)=-t_\q$ (we use $Z_{A,\Lambda}=1$) and $S_\Lambda=S+\Delta S_\Lambda$ corresponds to the local limit of decoupled sites (vanishing hopping amplitude), a limit which is exactly solvable. For small $k$, the function $R_k(\q)$ gives a mass $\sim k^2$ to the low-energy modes $|\q|\lesssim k$ and acts as an infrared regulator. 

The Bose-Hubbard model (with action $S_{k=0}$) can be related to the reference model (with action $S_\Lambda$) by a RG equation. We consider the scale-dependent effective action 
\begin{align}
\Gamma_k[\phi^*,\phi] ={}& - \ln Z_k[J^*,J] + \inttau \sum_\r (J^*_\r\phi_\r+\mbox{c.c.}) \nonumber \\ & - \Delta S_k[\phi^*,\phi] ,
\end{align}
defined as a (slightly modified) Legendre transform which includes the explicit subtraction of $\Delta S_k[\phi^*,\phi]$. Here $Z_k[J^*,J]$ is the partition function, $J_\r(\tau)$ a complex external source which couples linearly to the bosonic field and $\phi_\r(\tau)=\delta\ln Z_k[J^*,J]/\delta J^*_\r(\tau)$ is the superfluid order parameter. The variation of the effective action with $k$ is governed by Wetterich's equation~\cite{Wetterich93}, 
\beq
\partial_k \Gamma_k[\phi^*,\phi] = \half \Tr\biggl\lbrace \partial_k R_k\left(\Gamma^{(2)}_k[\phi^*,\phi] + R_k\right)^{-1} \biggr\rbrace ,
\label{floweq}
\eeq
where $\Gamma^{(2)}_k$ is the second-order functional derivative of $\Gamma_k$. In Fourier space, the trace in (\ref{floweq}) involves a sum over momenta and frequencies as well as the two components of the complex field $\phi$. The initial condition of the RG equation is 
\beq
\Gamma_\Lambda[\phi^*,\phi] = \Gamma_{\rm loc}[\phi^*,\phi] + \inttau \sum_\q \phi^*(\q) t_\q \phi(\q), 
\label{Gamin} 
\eeq
where $\Gamma_{\rm loc}[\phi^*,\phi]=- \ln Z_\Lambda[J^*,J] + \inttau \sum_\r (J^*_\r\phi_\r+\mbox{c.c.})$ is the Legendre transform of the free energy $- \ln Z_\Lambda[J^*,J]$ of the reference system corresponding to the local limit of decoupled sites. The effective action $\Gamma_\Lambda$ reproduces the strong-coupling RPA theory of the Bose-Hubbard model, which treats exactly the on-site repulsion but takes into account the inter-site hopping term in a mean-field type approximation~\cite{Sheshadri93,*Oosten01,Sengupta05,*Ohashi06,Menotti08}. The strong-coupling RPA theory describes qualitatively the phase diagram but is not quantitatively accurate and breaks down in the critical regime near the superfluid--Mott-insulator transition. In the NPRG technique, fluctuations beyond the RPA are included by solving the flow equation~(\ref{floweq}). Since the starting action $S_\Lambda$ is purely local, our approach is to some extent reminiscent of various $t/U$ expansions of the Bose-Hubbard model~\cite{Freericks96,*Buonsante05,*Koller06,*Freericks09,*Santos09,*Teichmann09a,*Knap10}. 

We are primarily interested in two quantities. The first one is the effective potential defined by $V_k(n)=\frac{1}{\beta N}\Gamma_k[\phi^*,\phi]$ with $\phi$ a constant (i.e. uniform and time-independent) field and $n=|\phi|^2$. Its minimum determines the condensate density $n_{0,k}$ and the thermodynamic potential (per site) $V_{0,k}=V_k(n_{0,k})$ in the equilibrium state. At the initial stage of the RG, $V_\Lambda(n)=V_{\rm loc}(n)-2dtn$, where $V_{\rm loc}(n)$ is the thermodynamic potential in the local limit. 

The second quantity of interest is the two-point vertex $\Gamma^{(2)}_k$ which determines the single-particle propagator $G_k=-\Gamma^{(2)-1}_k$ and therefore the excitation spectrum. Because of the U(1) symmetry of the action~(\ref{action}), the two-point vertex in a constant field takes the form 
\beq
\Gamma_{k,ij}^{(2)}(q;\phi) = \delta_{ij}\Gamma_{A,k}(q;n) + \phi_i\phi_j \Gamma_{B,k}(q;n) + \eps_{ij} \Gamma_{C,k}(q;n) 
\label{gam2}
\eeq
in Fourier space, where $q=(\q,i\w)$ and $\w$ is a Matsubara frequency. Here $(\phi_1,\phi_2)=\sqrt{2}(\mbox{Re}(\phi),\mbox{Im}(\phi))$, $n=|\phi|^2=\frac{1}{2}(\phi_1^2+\phi_2^2)$ and $\eps_{ij}$ is the antisymmetric tensor. In order to solve the flow equation~(\ref{floweq}), we use a derivative expansion of $\Gamma_k^{(2)}$, 
\beq
\begin{split} 
\Gamma_{A,k}(q;n) &= V_{A,k}(n) \w^2 + Z_{A,k}(n) \eps_\q + V_k'(n) , \\ 
\Gamma_{B,k}(q;n) &= V_k''(n) , \\ 
\Gamma_{C,k}(q;n) &= Z_{C,k}(n) \w ,
\end{split}
\label{gamde}
\eeq
where $\eps_\q=t_\q+2dt$ ($\eps_\q\simeq t\q^2$ for $|\q|\ll 1$). This derivative expansion is similar to the one used in continuum models~\cite{Dupuis07,*Wetterich08,*Floerchinger08,Dupuis09a,*Dupuis09b,Sinner09,*Sinner10}, but the initial conditions at scale $k=\Lambda$ are here obtained from $\Gamma^{(2)}_\Lambda$ [Eq.~(\ref{Gamin})] and therefore already include on-site quantum fluctuations. To reduce the numerical effort, one can further approximate $V_{A,k}(n)$ by $V_{A,k}\equiv V_{A,k}(n_{0,k})$ (and similarly for $Z_{A,k}(n)$ and $Z_{C,k}(n)$) and expand the effective potential to quadratic order about its minimum,
\beq
V_k(n) = \left\lbrace 
\begin{array}{lcc}
V_{0,k} + \frac{\lambda_k}{2}(n-n_{0,k})^2 & \mbox{if} & n_{0,k}>0 , \\ 
V_{0,k} + \Delta_k n + \frac{\lambda_k}{2}n^2 & \mbox{if} & n_{0,k}=0 .
\end{array}
\right. 
\label{trunc} 
\eeq
In the superfluid phase, Eqs.~(\ref{gamde},\ref{trunc}) yield a gapless mode $\w=c_k|\q|$ with velocity
\beq
c_k = \left(\frac{Z_{A,k}t}{V_{A,k}+Z_{C,k}^2/(2\lambda_k n_{0,k})}\right)^{1/2} 
\label{velocity}
\eeq
and a superfluid stiffness (defined as the rigidity wrt a twist of the phase of the order parameter) $\rho_{s,k}=2t Z_{A,k}n_{0,k}$. All physical quantities of interest can now be obtained by solving the flow equation~(\ref{floweq}) together with Eqs.~(\ref{gam2},\ref{gamde}) (and, possibly, Eq.~(\ref{trunc}))~\cite{note1}. In the following, we focus on the two-dimensional Bose-Hubbard model; the three-dimensional model will be discussed elsewhere. 
%
%
\begin{figure}
\centerline{\includegraphics[width=6.7cm,clip]{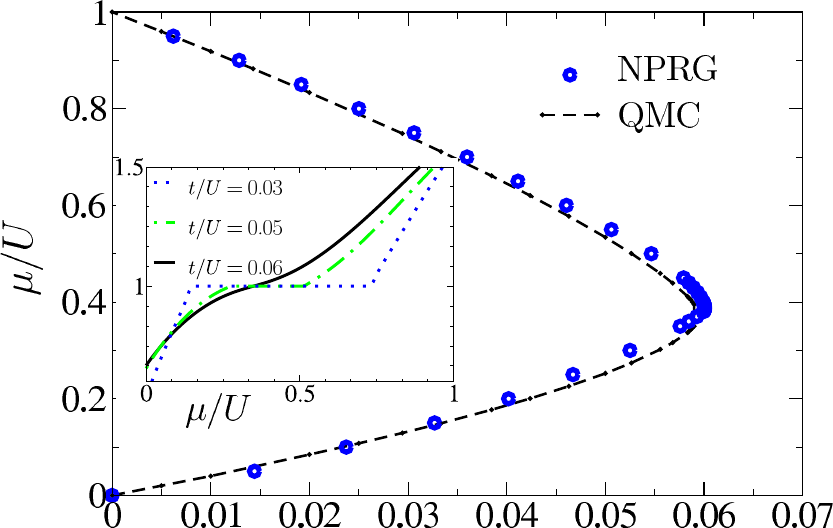}}
\vspace{0.3cm}
\centerline{\includegraphics[width=6.4cm,clip]{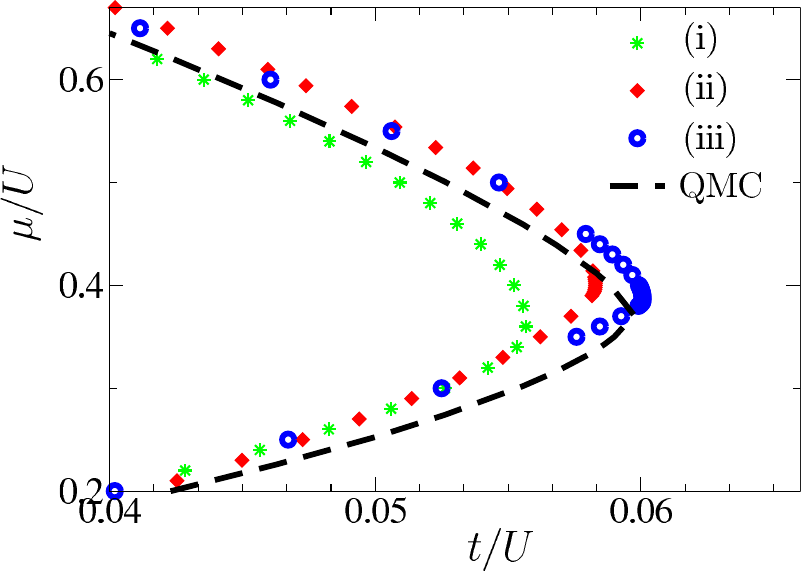}}
\caption{(Color online) Top panel: Phase diagram of the two-dimensional Bose-Hubbard model showing the first Mott lob with density $\bar n=1$. The NPRG result is shown by the (blue) points and the QMC data of Ref.~\cite{Capogrosso08} by the dashed line. Inset: density  $\bar n$ vs $\mu/U$ for different values of $t/U$. Bottom panel: Transition lines obtained from three different approximations (see text). The best one (iii) is also shown in the top figure.}
\label{phasediag}
\end{figure}
%
%

\textit{Phase diagram.} For given values of $t$, $U$ and $\mu$, the ground state can be deduced from the values of the condensate density  $n_{0,k=0}$ ($n_{0,k=0}>0$ in the superfluid phase), while the density is obtained from  $\bar{n}=-\frac{d }{d\mu}V_{0,k=0}$. It takes only a couple of seconds (depending on the approximation scheme, see below) to solve numerically the NPRG equations (for $t$, $U$ and $\mu$ fixed) on a standard PC, so that the full determination of the phase diagram requires at most an hour. 

Fig.~\ref{phasediag} shows the phase diagrams obtained from three different approximations: i) the effective potential $V_k(n)$ is truncated [Eq.~(\ref{trunc})] and the $n$ dependence of $V_{A,k}(n)$, $Z_{A,k}(n)$, $Z_{C,k}(n)$ is neglected as explained above; ii) the full $n$ dependence of $Z_{C,k}(n)$ is included; iii) the full $n$ dependence of $V_k(n)$ and $Z_{C,k}(n)$ is included. By including more functions into the analysis (i.e. going from (i) to (iii)) we observe a nice convergence of our results, which we therefore expect to be close to the exact ones, with a typical error, estimated from the difference between (ii) and (iii), roughly of order of $3\%$. This expectation is confirmed by a direct comparison to the QMC data~\cite{Capogrosso08}: the tip of the Mott lob ($t/U=0.060$, $\mu/U=0.387$) differs from the QMC result only by ($1.5\%$, $4\%$). It should be noted that the accuracy of the NPRG (within similar approximation schemes) in computing non-universal quantities (phase diagrams and thermodynamics) has been reported in other contexts, in particular in classical spin models~\cite{Machado10} and finite-temperature field theory~\cite{Blaizot11}. 

\begin{figure}
\centerline{\includegraphics[width=5.8cm,clip]{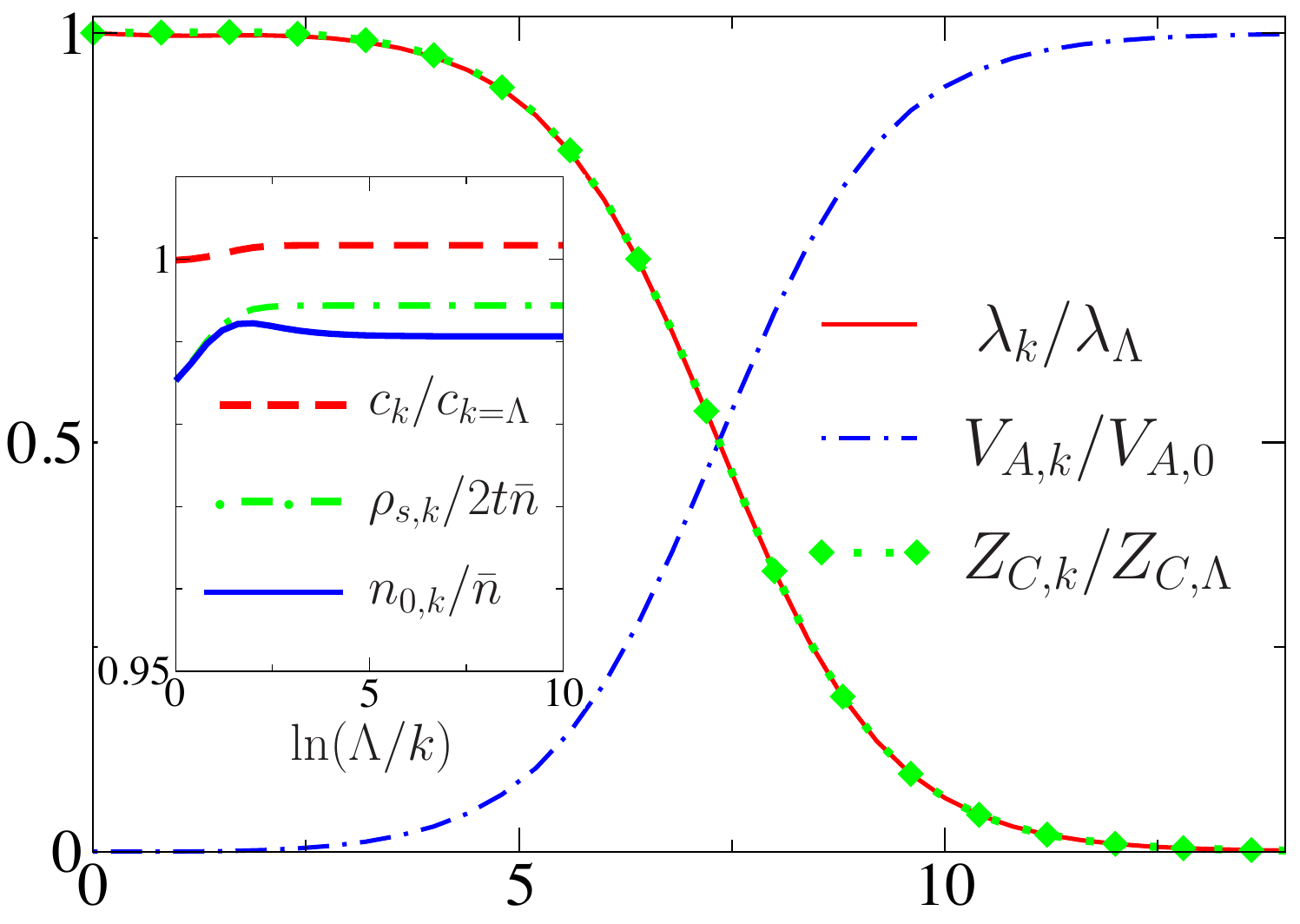}}
\centerline{\includegraphics[width=5.9cm,clip]{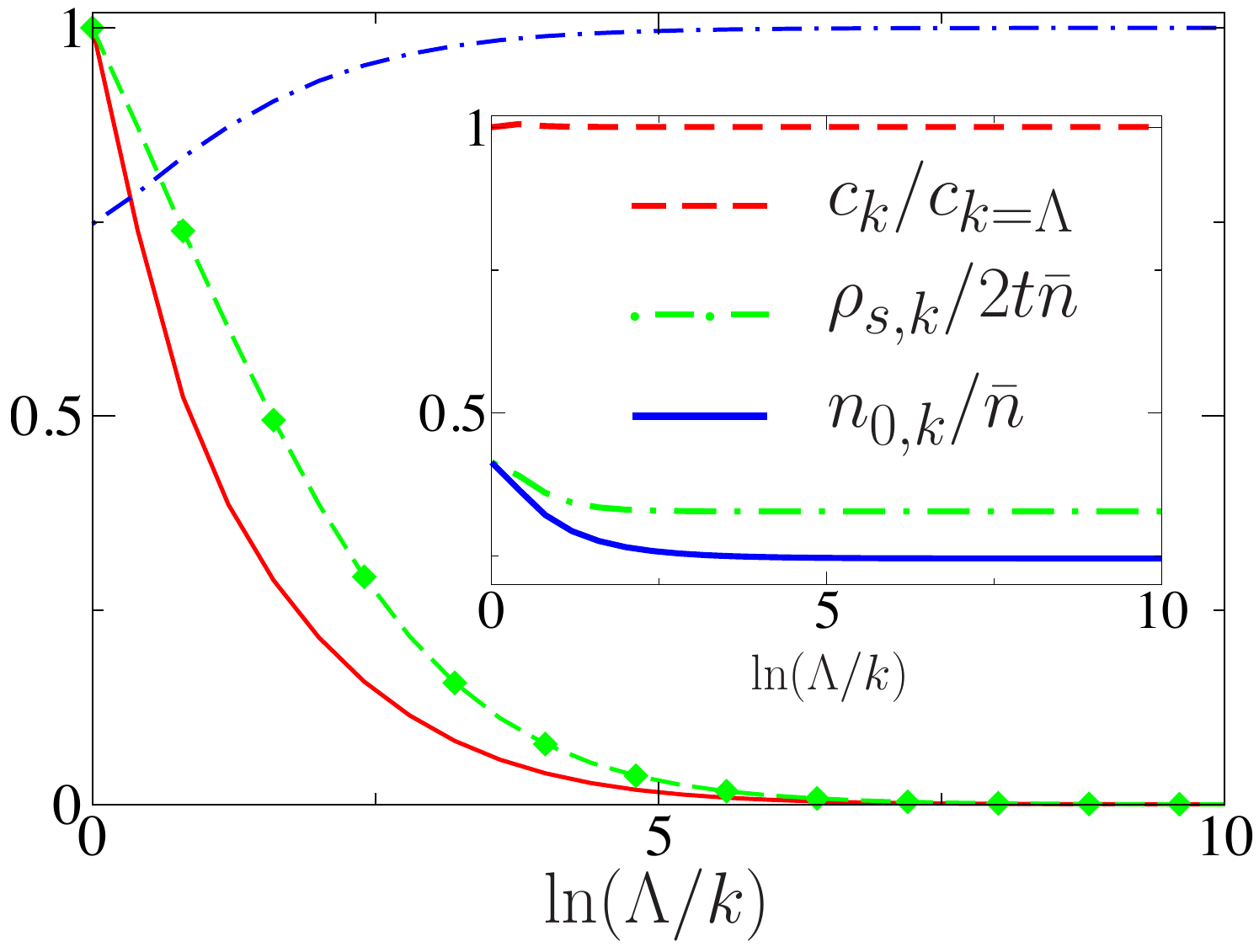}}
\caption{(Color online) NPRG flows in the weakly-correlated superfluid phase, $t/U=10$ and $k_G\ll k_h\ll \Lambda$ (top), and the strongly-correlated superfluid phase, $t/U\simeq 0.062$ and $k_h\sim k_G\sim \Lambda$ (bottom), for a density $\bar n=1$. The insets show $c_k$, $n_{0,k}$ and $\rho_{s,k}$ vs $\ln(\Lambda/k)$.}
\label{flow_SF}
\end{figure}

\textit{Superfluid phase.} 
In the weak-coupling limit, we recover the results of previous NPRG studies in continuum models~\cite{Dupuis07,*Wetterich08,*Floerchinger08,Dupuis09a,*Dupuis09b,Sinner09,*Sinner10}. The strong-coupling RPA (the initial condition of the RG flow) is equivalent to the Bogoliubov approximation when $U/t\ll1$~\cite{Menotti08}.  The condensate density $n_{0,k}$, the superfluid stiffness $\rho_{s,k}$ or the Goldstone mode velocity $c_k$ vary weakly with $k$ and are well approximated by their Bogoliubov estimates $n_{0,\Lambda}\simeq \bar n$, $\rho_{s,\Lambda}\simeq 2t\bar n$ and $c_{\Lambda}\simeq (2Ut\bar n)^{1/2}$ (Fig.~\ref{flow_SF}). On the other hand, from the strong variation of $\lambda_k$, $Z_{C,k}$ and $V_{A,k}$ with $k$, we can distinguish two regimes separated by the characteristic (Ginzburg) momentum scale $k_G=\xi_G^{-1}\sim \sqrt{\den (U/t)^{3}}$: i) a (perturbative) Bogoliubov regime $k\gg k_G$ where $\lambda_k \simeq \lambda_{\Lambda}\simeq U$, $Z_{C,k}\simeq Z_{C,\Lambda}\simeq 1$ and $V_{A,k}\simeq 0$. The spectrum crosses over from a quadratic dispersion to a linear sound-like dispersion at the (healing) momentum scale $k_h=\xi_h^{-1}\simeq\sqrt{\den U/t}$ defined by $n_{0,k}\lambda_k\simeq Z_{A,k}t k^2$. ii) a (non-perturbative) Goldstone regime $k\ll k_G$ where $\lambda_k$, $Z_{C,k}\sim k$ vanish with $k\to0$ and $V_{A,k}\simeq V_A^*$ takes a finite value. This regime is dominated by phase fluctuations, and characterized by the vanishing of the anomalous self-energy $\Sigma_{\text{an},k}(q=0)=\lambda_kn_{0,k}\sim k$ and the divergence of the longitudinal propagator $G_{\parallel,k}(q=0)=1/(2\lambda_kn_{0,k})$~\cite{Dupuis10}. A weakly-correlated two-dimensional superfluid is characterized by $k_G/k_h\sim U/t\ll1$; although the Bogoliubov theory breaks down at low energy, it applies to a large part ($k_G\lesssim |\vq|\lesssim k_h\ll\Lambda$) of the spectrum where the dispersion is linear.

\begin{figure}
\centerline{\includegraphics[width=6.cm,clip]{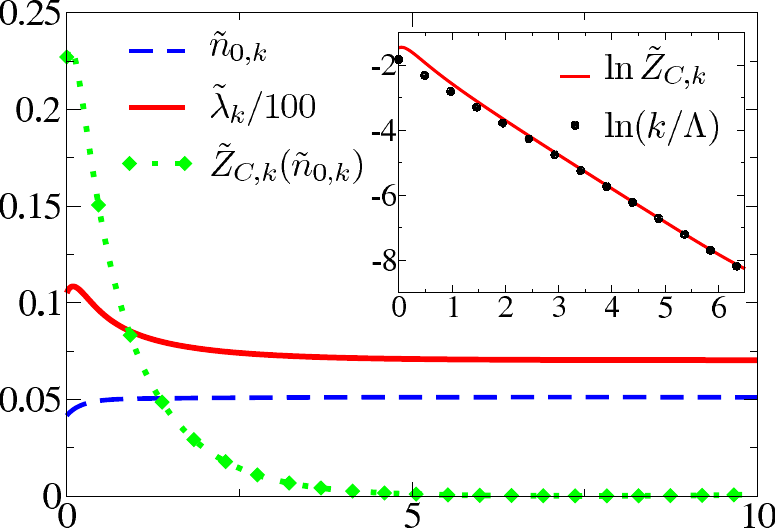}}
\vspace{0.15cm}
\centerline{\includegraphics[width=6.cm,clip]{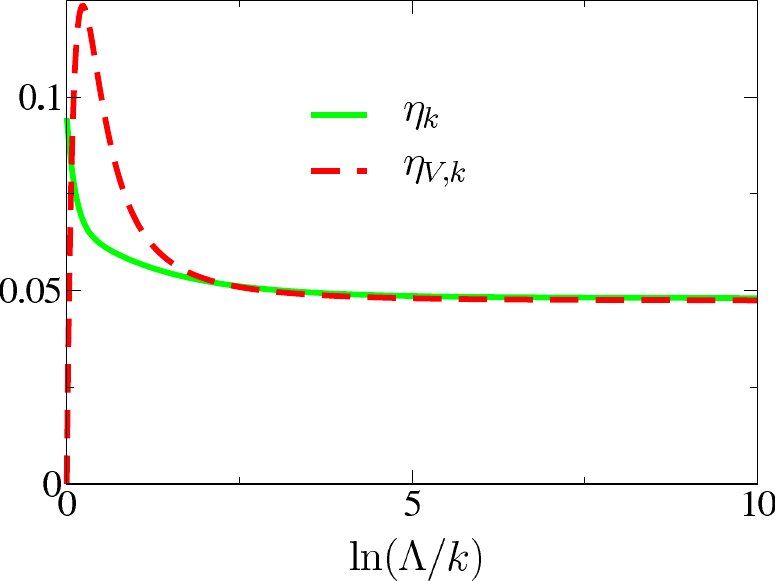}}
\caption{(Color online) Top panel: Dimensionless condensate density $\tilde{n}_{0,k}$ and coupling constant $\tilde{\lambda}_k$ vs $\ln(\Lambda/k)$ at the XY critical point (the inset shows $\ln \tilde Z_{C,k}(\tilde n_{0,k})$ vs $\ln(\Lambda/k)$). Bottom panel: (running) anomalous dimensions $\eta_{k}=-k\partial_k \ln Z_{A,k}(n_{0,k})$ and $\eta_{V,k}=-k\partial_k \ln V_{A,k}(n_{0,k})$. The dynamical exponent is defined by $z_k=(2-\eta_k+\eta_{V,k})/2$.}
\label{flowXY}
\end{figure}

As $t/U$ decreases, the ratio $k_G/k_h$ increases and eventually becomes of order one (with $k_h\sim k_G\sim \Lambda$). In this strongly-correlated superfluid phase, there is no Bogoliubov regime any more and the condensate density $n_0\equiv n_{0,k=0}$, as well as the superfluid stiffness $\rho_s\equiv\rho_{s,k=0}$, is strongly suppressed (Fig.~\ref{flow_SF}). 

{\it Critical regime.} Our approach recovers the two universality classes of the superfluid--Mott-insulator transition~\cite{Fisher89}. Away from the tip of the Mott lob, the transition is mean-field like (with logarithmic corrections) with a dynamical exponent $z=2$. At the tip, it belongs to the universality class of the three-dimensional XY model. Fig.~\ref{flowXY} shows the RG flows of the dimensionless coupling constants 
\begin{equation}
\begin{split} 
\tilde{n}_{0,k} &=k^{-d}\left(Z_{A,k} t k^{2}V_{A,k}\right)^{1/2} n_{0,k} , \\ 
\tilde{\lambda}_k &= k^d V_{A,k}^{-1/2}\left(Z_{A,k} t k^2\right)^{-3/2} \lambda_k ,\\
\tilde{Z}_{C,k}(\tilde{n}_{0,k}) &= \left(V_{A,k}Z_{A,k} t k^2\right)^{-1/2}Z_{C,k}(n_{0,k}),
\end{split}
\end{equation}
when the system is at the XY critical point. The plateaus observed for the dimensionless condensate density $\tilde n_{0,k}$ and coupling constant $\tilde\lambda_k$, as well as for the (running) anomalous dimensions $\eta_k=-k\partial_k \ln Z_{A,k}(n_{0,k})$ and $\eta_{V,k}=-k\partial_k \ln V_{A,k}(n_{0,k})$, are characteristic of critical behavior. We find the critical exponent $\nu=0.699$, the anomalous dimensions $\eta=0.049$, $\eta_V=\eta(1-\eta/4)= 0.049$, and the dynamical exponent $z=(2-\eta+\eta_{V})/2=1.000$, to be compared with the best known estimates $\nu=0.671$ and $\eta=0.038$ for the three-dimensional XY model~\cite{Campostrini01}. Table~\ref{table} summarizes the infrared behavior of the two-dimensional Bose-Hubbard model. Note that both in the superfluid phase and at the XY critical point, the infrared behavior is characterized by a relativistic symmetry ($Z_{C,k}\to 0$ for $k\to 0$). 

\begin{table}
\renewcommand{\arraystretch}{1.5}
\begin{center}
\begin{tabular}{|c|c||c|c|c|c|c|}
\hline 
\multicolumn{2}{|c||}{} & $Z_{A,k}$ & $V_{A,k}$ & $Z_{C,k}$ & $\lambda_k$ & $n_{0,k}$
\\ \hline \hline 
\multicolumn{2}{|c||}{superfluid} & $Z_A^*$ & $V_A^*$ & $k$ & $k$ & $n_0^*$ 
\\ \hline 
critical & $XY$ & $k^{-\eta}$ & $k^{-\eta}$ & $k$ & $k^{1-2\eta}$ & $k^{1+\eta}$ 
\\ \cline{2-7} 
behavior & mean-field & $Z_A^*$ & $V_A^*$ & $Z_C^*$ & $|\ln k|^{-1}$ & $k^2|\ln k|^{-1}$ 
\\ \hline 
\multicolumn{2}{|c||}{insulator} & $Z_A^*$ & $V_A^*$ & $Z_C^*$ & $\lambda^*$ & 0 
\\ \hline 
\end{tabular}
\end{center}
\caption{Critical behavior at the superfluid--Mott-insulator transition and infrared behavior in the superfluid and Mott-insulator phases. The stared quantities indicate nonzero fixed-point values and $\eta$ denotes the anomalous dimension at the three-dimensional $XY$ critical point. $Z_{A,k}$ and $V_{A,k}$ stand for $Z_{A,k}(n_{0,k})$ and $V_{A,k}(n_{0,k})$.}
\label{table}
\end{table}

\textit{Conclusion.} The excellent agreement between our results and the QMC data~\cite{Capogrosso08} shows that the lattice NPRG, first introduced in Ref.~\cite{Machado10} for classical systems, is a very efficient method to study the Bose-Hubbard model. This RG approach, which is implemented in momentum space, takes into account local fluctuations and Mott physics while being able to describe critical fluctuations.

We would like to thank B. Delamotte for useful discussions and B. Capogrosso-Sansone for providing us with the QMC data shown in Fig.~\ref{phasediag}.

\vspace{-0.25cm} 


\begin{thebibliography}{28}
\bibitem{note1} Note that with the parametrizations (\ref{gam2}), (\ref{gamde}) and (\ref{trunc}), the flow equations are the same as those derived in Ref.~\cite{Dupuis09b} (see, in particular, Appendix C in the second paper) but with different initial conditions.
\end{thebibliography}
%
\end{document}